\documentclass[thmsa,12pt]{article}

\usepackage{amssymb}
\usepackage{float}
\usepackage{makeidx}
\usepackage[dvips]{graphicx}
\usepackage{indentfirst}

\title{\bf Conductivity and Viscosity Measurements for Binary Lysozyme Chloride
Aqueous Solution and Ternary Lysozyme-Salt-Water Solution}

\author{Daniela Buzatu\thanks{Physics Department,
Politehnica University, Bucharest, 77206, Romania.}, Emil
Petrescu\footnotemark[1],
 Florin D. Buzatu\thanks{Department of
Theoretical Physics, National Institute for Physics and Nuclear
Engineering, Bucharest-M\u{a}gurele, 76900, Romania.}}

\date{}
\begin{document}

\maketitle

\setlength{\unitlength}{1mm}
\newcommand{\pl}{\partial}
\newcommand{\ep}{\varepsilon}

\begin{abstract}

\noindent We use the conductimetric method, adequate to
electrolytes, to determine the lysozyme charge in lys-water and
ternary lys-salt-water systems. We measured also the viscosities
for the above binary and ternary systems in the same conditions at
pH$=4.5$ and T$=298$ K, measurements that allow us to see any
effect of viscosity on cations mobilities and implicitly on the
lysozyme charge. The method is illustrated for the lysozyme
chloride aqueous solution system at 25$^o$ C, using the data
reported here for pH$=4.5$ at 0.15, 0.6, 0.8, 1., 1.5, 2., 2.5,
3., 3.5 mM (mg/mL) lysozyme chloride concentrations. The method
was also applied to ternary lys-salt-water systems in the same
conditions at pH$=4.5$ and T$=25^o$ C. Ternary conductivities are
reported for a mean concentration 0.6 mM  of lysozyme chloride in
all systems and a mean concentration 0.01, 0.025, 0.05, 0.1,
0.175, 0.2, 0.5, 0.7, 0.9, 1.2, 1.3 and 1.4 M for NaCl; 0.005,
0.01, 0.05, 0.1, 0.175, 0.2, 0.5, 0.7, 0.9, 1.2, 1.3, 1.4 and 1.5
M for KCl; 0.005, 0.01, 0.05, 0.1, 0.175, 0.2, 0.5, 0.7, 1.2, 1.3,
1.4, 1.5, 1.6 and 1.7 M for NH$_4$Cl.

\end{abstract}

\newpage
\noindent {\bf Motivation} \\

\noindent In order to determine the protein structure through
X-ray diffraction, high quality crystals are required. The protein
crystallization process usually occurs in aqueous solution that
contain salts as precipitant agents. The physical and chemical
properties of these solutions affect drastically the nucleation
and crystal growth processes. Protein aggregation depends on
protein-protein and protein-precipitant interactions in the
solution. In these interactions, the effective charge of the
protein plays an important role. From titration experiments
\cite{1} only the stoichiometric value of the protein charge can
be determine, but it does not take into account the presence of
ions that may bind on the macromolecules and change their net
charge as already shown by diffusion experiments \cite{2}. \\
\noindent A more direct way to estimate the protein charge for
different salts concentrations is based on conductimetric
experiments. The conductivity \cite{16,17} of the protein-salt
aqueous systems is function of the mobility and of the ionic
charge species present in solution and it seems to be very
sensitive to charge changes of the protein. These changes are
influenced by
modifications in pH, salt concentrations and the type of the used salt. \\

\noindent {\bf Theoretical Background} \\

\noindent Assuming that the lysozyme chloride is an electrolyte
with the chemical formula Lys$^{z_p}$Cl$^-_{z_p}$, the
conductivity for a ternary system
Lys$^{z_p}$Cl$^-_{z_p}$-Salt-Water can be expressed as:
\begin{eqnarray}
\sigma_{sol}=c_pF(z_pu_p+z_pu_-)+c_{salt}F(u_++u_-)
\end{eqnarray}
where $c_p$ is the protein concentration, F$=96484$ C/mol is the
Faraday constant, $u_p$, $u_-$ and $u_+$ are the mobilities of the
protein complex ion, the negative and positive ion of the used
salt in the solution, respectively. Using the Stoke's law and the
definition of the ion mobility, we can express the ion mobility in
terms of the viscosity of the medium and of the radius $r_i$ of
the ion in the solution (by considering the ion a hard sphere). We
assume also the radius of the ions doesn't change with the salt
concentration at very low protein concentrations (0.6 mM in our case).\\
\noindent Using the Einstein's equation for mobility of protein
complex ion at infinite dilution
\begin{eqnarray}
u_p^{\infty}=\frac{Fz_p^{\infty}D_p^{\infty}}{RT}\;,
\end{eqnarray}
where R$=8310$ J/mol K is the universal gas constant and
$D_p^{\infty}=0.132\times 10^{-9}$ m$^2$/s is the infinite
dilution tracer diffusion coefficient for the lysozyme \cite{2},
and Stokes-Eistein's equation
\begin{eqnarray}
D_p^{\infty}=\frac{kT}{6\pi\eta_Wr_p}\;,
\end{eqnarray}
where k-Boltzmann's constant, and $\eta_W=0.8904$ cp is the water
viscosity at T$=298$ K, we can write the conductivity of the
ternary solution as:
\begin{eqnarray}
\sigma_{sol}=c_p\Lambda_{p}+c_{salt}\Lambda_{salt}
\end{eqnarray}
where
\begin{eqnarray}
\Lambda_p & = & z_pu_p+z_pu_-=z_p^2\lambda_p+z_p\lambda_- \\
\Lambda_{salt} & = & u_++u_-=\lambda_++\lambda_-
\end{eqnarray}
are the conductances for lysozyme chloride (protein) and
respectively the conductance for used salt. The specific
conductance of the positive/negative ions $\lambda_{\pm}$ and of
the protein ion can be determine from:
\begin{eqnarray}
\lambda_{\pm}=\frac{\eta_{salt}}{\eta_{sol}}t_{\pm}\Lambda_2^{\infty}
\end{eqnarray}
\begin{eqnarray}
\lambda_p=\frac{F^2}{RT}\frac{\eta_{water}}{\eta_{sol}}D_p^{\infty}
\end{eqnarray}
where $t_{\pm}$ is the transference number of the
positive/negative ions of the used salt in the binary solution
\cite{3}, $\eta_{sol}$ is the viscosity of the ternary solution,
$\eta_{salt}$ is the viscosity of the binary salt solution, and
$\Lambda_{salt}^{\infty}$ is the molar conductance of the salt
solution at infinite dilution. \\
\noindent From the Eqs. (5)-(8), we obtain the following
expression for $\sigma_{sol}$:
\begin{eqnarray}
\sigma_{sol}=c_p\left(z_p^2\frac{F^2}{RT}\frac{\eta_{water}}{\eta_{sol}}D_p^{\infty}+
z_p\frac{\eta_{salt}}{\eta_{sol}}t_-\right)+c_{salt}\frac{\eta_{salt}}{\eta_{sol}}\Lambda_{salt}^{\infty}
\nonumber \\
-c_p\left(z_p-z_p^o\right)\frac{\eta_{salt}}{\eta_{sol}}t_+\Lambda_{salt}^{\infty}
\end{eqnarray}
where z$_p^o=6.7$ is the lysozyme charge at infinite dilution
\cite{2}.

\noindent The Eq. (9) shows the strong dependence of the
($\sigma_{sol}$) on the protein charge, z$_p$. \\

\noindent {\bf Experimental section} \\

\noindent All the experimental work was performed at Texas
Christian University, Chemistry Department. \\ \\
\noindent {\bf Materials}. The materials, solution preparation,
density and pH measurements are all described in the Ref.
\cite{2}. We used a hen egg-white lysozyme, recrystallized six
times, purchased from Seikagaku America. The choice of the
supplier was guided by the work of Rosenberger and co-workers
\cite{4,5,6} who reported detailed analysis of commercial HEWL
products. \\
\noindent The molecular mass of the isoelectric lysozyme solute,
$M_{lys}$, was taken as 14307 g/mol \cite{7}, and this value was
used to calculate all concentrations after correction for the
moisture and chloride content. Buoyancy corrections were made with
the commonly used lysozyme crystal density \cite{8,9,10} of 1.305 g/cm$^3$.\\
Deionized water was distilled and then passed through a four-stage
Millipore filter system to provide high-purity water for all the
experiments. The molecular mass of water, $M_{water}$, was taken
as 18.015 g/cm$^3$. Mallinckrodt reagent HCl ($\sim$ 12 M) was
diluted by half with pure water and distilled at the constant
boiling composition. The resulting HCl solution ($\sim $ 6 M) was
then diluted to about 0.063 M (pH 1.2) and used to adjust the
pH of the solution.\\
\noindent Following the work of Rard \cite{11}, Mallinckrodt AR
NaCl and KCl were dried by heating at 450$^o$ C for 7 h, and used
without further purification. Mallinckrodt AR NH$_4$Cl was dried
heating it at 70$^o$ C for 7 h under vacuum. The purity of the
salts were listed as 99.9\% by the supplier. The molecular mass
for NaCl, KCl and NH$_4$Cl, M$_{salt}$, were taken to be
respectively 58.443 g/mol, 74.55 g/mol and 53.49 g/mol, and their
crystal density \cite{12} as 2.165 g/cm$^3$, 1.984 g/cm$^3$ and
1.527 g/cm$^3$ for buoyancy corrections. \\ \\
{\bf Preparation of Solutions}. All solutions were prepared by
mass with appropriate buoyancy corrections. All weighings were
performed with a Mettler Toledo AT400 electrobalance. Since the
as-received lysozyme powder was very hygroscopic, all
manipulations in which water absorption might be critical were
performed in a dry glove box. Stock solutions of lysozyme were
made by adding as-received protein to a pre-weighted bottle that
had contained dry box air, capping the bottle, and reweighing to
get the weight and thus mass of lysozyme. Water was added to
dissolve the lysozyme, and the solution was weighed. An accurate
density measurement was made and used to obtain the molarity of
the stock
solution. \\
For binary experiments, the solutions were first diluted to within
10 cm$^3$ of the final volumes with pure water. From 1 to 3 mL  of
the dilute HCl was added to adjust the pH to the desired value,
any residual solution on the pH electrode was washed back into the
solutions, and the dilutions were completed by mass. The densities
and final pH values of these solutions were measured and the final
concentrations were calculated. \\
For ternary experiments, precise masses of NaCl, KCl and NH$_4$Cl
were added to flasks containing previously weighed quantities of
lysozyme stock solutions. These solutions were mixed and diluted
to within 10 cm$^3$ of the final volume. The pH was adjusted, and
the solutions were diluted to their final mass. \\ \\
{\bf pH Measurements}. The pH measurements were done using a
Corning model 130 pH meter with an Orion model 8102 combination
ROSS pH electrode. The meter was calibrated with standard pH 7 and
pH 4 buffers and checked against a pH 5 standard buffer. It was
assumed that the pH values remaind valid at higher NaCl, KCl and
NH$_4$Cl concentrations. After four or five experiments, the
electrode was soaked in 5$\%$ NaClO for 10 min, and then the
internal reference solution was replaced with fresh solution.
\\ \\
{\bf Density Measurements}. All density measurements were
performed with a Mettler-Paar DMA40 density meter, with an RS-232
output to a Apple $\Pi$+. By time averaging the output, a
precision of 0.00001 g/cm$^3$ or better could be achieved. The
temperature of the vibrating tube in the density meter was
controlled with water from a large well-regulated water bath whose
temperature was 25.00$\pm$ 0.01 $^o$ C.
 \\ \\
{\bf Conductivity and Viscosity Measurements}. We performed
measurements on conductivity for ternary systems: Lys-NaCl-Water,
Lys-KCl-Water and Lys-NH$_4$Cl-Water, and for binary Lys-Water
using a pair of capillarity cells characterized by the length (l)
and cross section (S), and 8 silver-chloride electrodes; 2 current
carrying electrodes, 2 bridge electrodes between cells and 4 probe electrodes. \\
The DC-current intensity (I) and the drop tension (U) were
determined using a AMEL-Instruments galvanostat for very low
current intensity ($\mu$A) and a Hewlett Packard multimeter which
allowed us to measure very high resistance (M$\Omega$)
corresponding to very low concentration of the studied solutions
(mM). The internal resistance of multimeter was adjusted to a
value bigger than 10 G$\Omega$. Thus, the measurements of the drop
tensions on systems with a resistance of 10 M$\Omega$ are not
affected by
significant errors. \\
In order to determine the geometrical factor of the cells, we
calibrated them with KCl (very high purity) solution for a large
range of the concentration (0.5 mM up to 1.5 M). \\
The binary and ternary protein solutions, and the binary
salt-water systems were prepared as we described above, at the
same pH=4.5; conductivities and viscosities were measured at
T$=25^o$ C with a very good accuracy. \\
Ternary conductivities are reported for a mean concentration 0.6
mM  of lysozyme chloride in all systems and a mean concentration
0.01, 0.025, 0.05, 0.1, 0.175, 0.2, 0.5, 0.7, 0.9, 1.2, 1.3 and
1.4 M for NaCl; 0.005, 0.01, 0.05, 0.1, 0.175, 0.2, 0.5, 0.7, 0.9,
1.2, 1.3, 1.4 and 1.5 M for KCl; 0.005, 0.01, 0.05, 0.1, 0.175,
0.2, 0.5, 0.7, 1.2, 1.3, 1.4, 1.5, 1.6 and 1.7 M for NH$_4$Cl. \\
Binary conductivities for Lysozyme chloride-Water are reported for
a mean concentration 0.15, 0.6, 0.8, 1., 1.5, 2., 2.5, 3., 3.5 mM
of lysozyme chloride.\\
Binary conductivities for Salt-Water are reported for the same
concentrations used in the ternary solutions. \\
For the same binary and ternary systems we performed measurements
of viscosity using an Ostwald viscosimeter. \\

\noindent {\bf Results} \\

\noindent {\bf Binary lysozyme-water conductivity}. Conductivities
were determined from Eq. (10) the in the concentration range from
0.15 mM to 3.5 mM using the experimental data for DC-current
intensity (I), the drop tension (U) on the cells and the diameter
of the all conductivity cells $d=2$ mm:
\begin{eqnarray}
\sigma_p=\frac{I\cdot l}{S\cdot U}
\end{eqnarray}
After that we made the correction due to the geometrical factor.
The obtained values are plotted in Fig.1, where the fitting curve
can be expressed by the following formula:
\begin{eqnarray} \sigma_p=(0.0040\pm 0.0006)+c_p[(90.71\pm
1.39)-(371.7\pm 23.1)\sqrt{c_p}]
\end{eqnarray}
with $\chi^2=4.6\times 10^{-6}$. Using the fitting Eq. (11), we
expressed the conductance $\Lambda_p$ as:
\begin{eqnarray}
\Lambda_p=(90.71\pm 1.39)-(371.7\pm 23.1)\sqrt{c_p}
\end{eqnarray}
The first term in Eq. (12) represents the lysozyme chloride
conductance at infinite dilution ($c_p\rightarrow 0$),
$\Lambda_p^{\infty}=0.091$ Sm$^2$/mol. Taking into account the Eq.
(9), we can write the lysozyme chloride conductance at infinite
dilution as:
\begin{eqnarray}
\Lambda_p^{\infty}=\frac{F^2}{RT}\left((z^o_p)^2D_p^{\infty}+z^o_pD_{Cl}^{\infty}\right)
\end{eqnarray}
where $D_p^{\infty}=0.132\times 10^{-9}$ m$^2$/s is the infinite
dilution tracer diffusion coefficient for the lysozyme \cite{2}
and $D_{Cl}^{\infty}=2.03\times 10^{-9}$ m$^2$/s is the infinite
dilution tracer diffusion coefficient for chloride ion obtained
from the limiting ionic conductances \cite{13}. \\
Using the Eq. (13) and the data from conductimetric experiments,
we calculated the lysozyme charge, z$^o_p=7.9$. This value can be
compare with the lysozyme charge calculated from extrapolated
limiting diffusion coefficients, in binary lysozyme-water
diffusion experiments \cite{2}, z$^o_p=6.7$ at
pH$=4.5$ and T$=298$ K. \\

\begin{figure}
\hspace{2.7cm}
\includegraphics[scale=1]{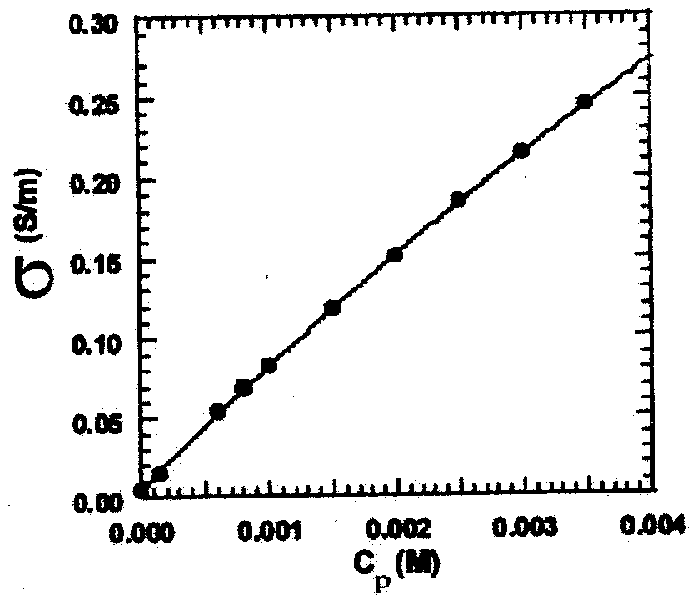}
\begin{center}
{\bf Fig. 1}
\end{center}
\end{figure}

\noindent {\bf Ternary lys-salt-water conductivity}. Using the
same Eq. (10) for ternary lys-salt-water solution, we determined
the conductivities for Lys-NaCl-Water, Lys-KCL-Water and
Lys-NH$_4$-Water. For these three ternary systems, we plotted in
Fig. 2 the experimental data for the difference
$\sigma_{sol}-\sigma_{salt}$ as function of salt concentration
$c_{salt}$. The used fitting curves are given bellow:
\begin{eqnarray}
\sigma_{lys-NaCl-w}-\sigma_{NaCl} & = & (0.0450\pm
0.0038)-(0.1625\pm
0.0053)c_{NaCl} \\
\sigma_{lys-KCl-w}-\sigma_{KCl}& = & (0.0281\pm
0.0053)-(0.1174\pm 0.0066)c_{KCl} \\
\sigma_{lys-NH_4Cl-w}-\sigma_{NH_4Cl} & = & (0.0401\pm
0.0052)-(0.1215\pm 0.0055)c_{NH_4Cl}
\end{eqnarray}

\noindent In the Fig. 3 we plotted the experimental data measured
for binary NaCl-Water, KCl-Water and NH$_4$-Water viscosities.
Below are given the fitting curves:
\begin{eqnarray}
\eta_{NaCl} =  0.8904+(0.0736\pm 0.0041)c_{NaCl} \nonumber \\
+(0.0043\pm 0.0089)c^2_{NaCl} +(0.0043\pm 0.0089)c^3_{NaCl}
\end{eqnarray}
\begin{eqnarray}
\eta_{KCl}  =  0.8904-(0.0025\pm 0.0035)c_{KCl}\nonumber \\
-(0.0054\pm 0.0069)c^2_{KCl} +(0.0055\pm 0.0033)c^3_{KCl}
\end{eqnarray}
\begin{eqnarray}
\eta_{NH_4Cl}  =  0.8904-(0.0098\pm 0.0037)c_{NH_4Cl}\nonumber \\
+(0.0045\pm 0.0064)c^2_{NH_4Cl}-(0.0009\pm 0.0027)c^3_{NH_4Cl}
\end{eqnarray}

\begin{figure}
\hspace{2cm}
\includegraphics[scale=1]{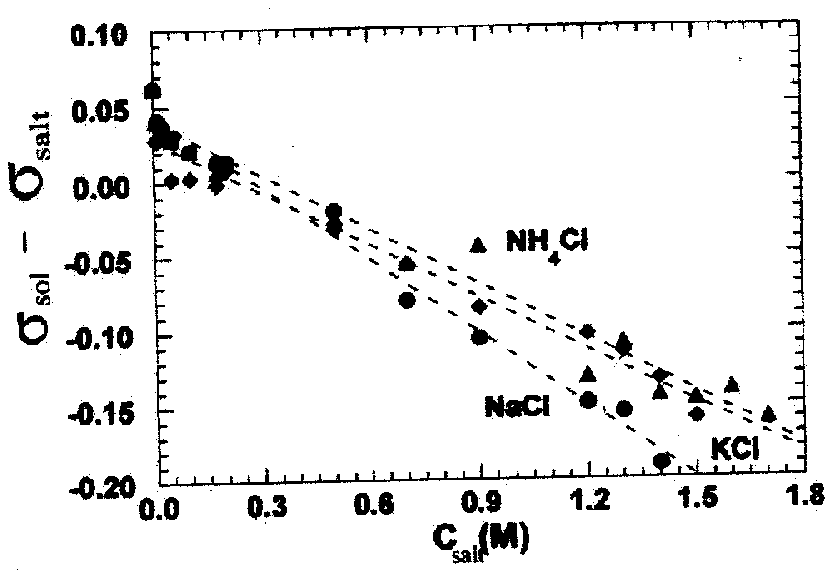}
\begin{center}
{\bf Fig. 2}
\end{center}
\end{figure}

\noindent In Fig. 4 is plotted the ratio $\eta{salt}/\eta_{sol}$
and the lysozyme effect on viscosity can be observed. For
$c_{salt}>0.05$ M the lysozyme has a constant effect on all the
three salts.

\begin{figure}
\hspace{2.2cm}
\includegraphics[scale=1]{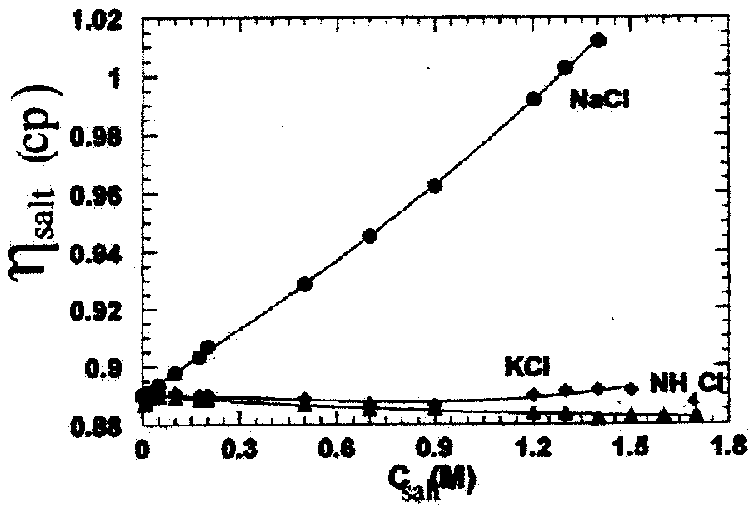}
\begin{center}
{\bf Fig. 3}
\end{center}
\end{figure}

\noindent Using the Eq. (9), we calculated also the lysozyme
charge dependence on salt concentrations; the results are shown in
Table 1.
\begin{table}[h]
\begin{center}
{\bf Table 1} \\ \vspace{0.3cm}
\begin{tabular}{ccccccccc}
 $c_{NaCl}$(M) & $z_p$ & $\delta z_p$ & $c_{KCl}$(M) & $z_p$ &
 $\delta z_p$ & $c_{NH_4Cl}$(M) & $z_p$ & $\delta z_p$ \\
 & & & & & & & & \\
 0  & 7.12 & 1.8 & 0.05 & 6.42 & 2.6 &  0 & 5.96 & 2.8 \\
 0.01 & 7.65 & 1.7 & 0.1 & 8.62 & 1.9 & 0.005 & 6.57 & 2.5 \\
 0.025 & 8.07 & 1.6 & 0.175 & 11 & 1.5 & 0.01 & 7.02 & 2.4 \\
 0.05 & 8.64 & 1.6 &  0.2 & 11.6 & 1.5 & 0.05 & 8.98 & 1.9 \\
 0.1 & 9.5 & 1.5 & 0.5 & 17.1 & 1 & 0.1 & 10.6 & 1.6 \\
 0.175 & 10.5 & 1.3 & 0.7 & 19.8 & 0.9 & 0.175 & 12.5 & 1.3 \\
 0.2 & 10.8 & 1.3 & 0.9 & 22 & 0.8 & 0.2 & 13.1 & 1.3 \\
 0.5 & 13.3 & 1.1 & 1.2 & 24.9 & 0.7& 0.5 & 18.1 & 0.9 \\
 0.7 & 14.4 & 1.1 & 1.3 & 25.8 & 0.7 & 0.7 & 20.5 & 0.8 \\
 0.9 & 15.1 & 1.1 & 1.4 & 26.6 & 0.6 & 0.9 & 22.5 & 0.8 \\
 1.2 & 15.9 & 1 & 1.5 & 27.4 & 0.6 & 1.2 & 25.2 & 0.7 \\
 1.3 & 16.1 & 1 &  & & & 1.3 & 26 & 0.7 \\
 1.4 & 16.3 & 1 & & & & 1.4 & 26.8 & 0.6 \\
 & & & & & & 1.5 & 27.4 & 0.6 \\
  & & & & & & 1.6 & 28.2 & 0.6 \\
  & & & & & & 1.7 & 28.8 & 0.6
\end{tabular}
 \end{center}
\end{table}

\noindent The propagation of errors was done assuming an error
$\delta\sigma=0.01$ S/m estimated by the least square on raw data.
We do not consider any systematic error associated to the model or
to the binary salt conductivities.

\begin{figure}
\hspace{2.5cm}
\includegraphics[scale=1]{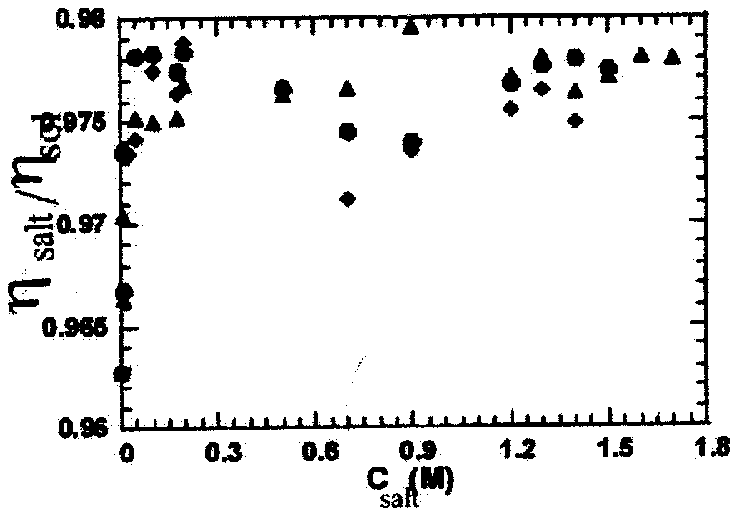}
\begin{center}
{\bf Fig. 4}
\end{center}
\end{figure}

\noindent In the Fig. 5 are plotted the values obtained for
lysozyme charge versus the salt concentration. The graph shows a
higher dependence of the lysozyme charge on KCl and NH$_4$Cl salt
concentrations than in the  NaCl case; it follows z$_p$ depends on
type of salt.

\begin{figure}
\hspace{1.7cm}
\includegraphics[scale=1]{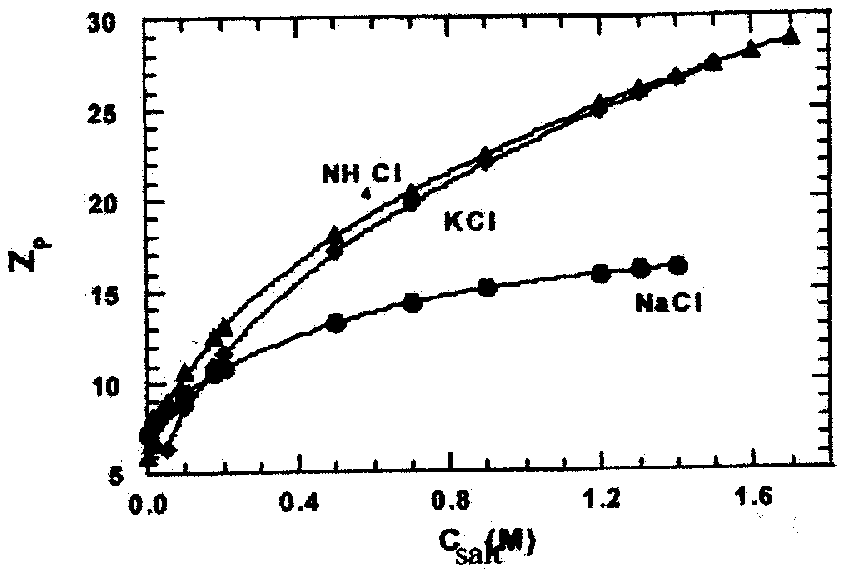}
\begin{center}
{\bf Fig. 5}
\end{center}
\end{figure}

\noindent This behavior may be connected with the dependence of
the mobility ratio Lys/Cation ($u_p/u_+$) on the salt
concentrations as it is shown in Fig. 6.

\begin{figure}
\hspace{1.5cm}
\includegraphics[scale=1]{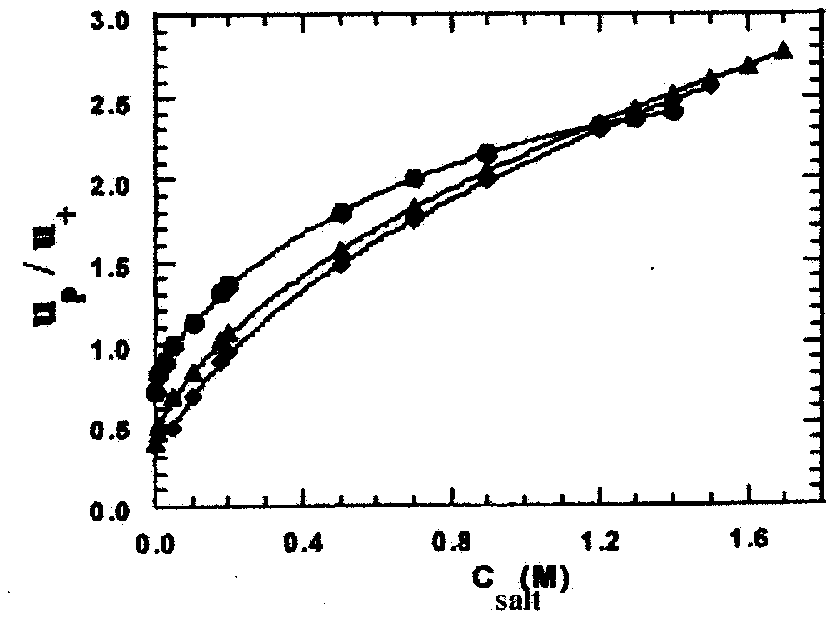}
\begin{center}
{\bf Fig. 6}
\end{center}
\end{figure}

\noindent The lysozyme charge values in the case when the salt
concentration goes to zero in ternary systems, are: $z_p=7.12$ for
ternary Lys-NaCl-Water solution, $z_p<6.42$ for ternary
Lys-KCl-Water solution and $z_p=5.95 $ for ternary
Lys-NH$_4$Cl-Water solution. We can compare the value of $z_p$
obtained for Lys-NaCl-Water by conductimetric method with the
value $z_p=8.9$ obtained from thermodynamic data applied to
precision ternary diffusion data for the same system \cite{14}. \\

\noindent {\bf Conclusions} \\

\noindent In the present paper we calculated the protein charge by
using, for the first time in our knowledge, a conductimetric
method to binary lysozyme-water, and ternary lys-salt-water
systems for different salts, at different salt concentrations. We
used also the experimental values of viscosities for the same
binary and ternary systems at pH$=4.5$ and T$=298$ K. We
determined the protein charge z$_p$ from Eq. (9) where we
introduced all the experimental data. \\
From binary lys-water conductivities data we obtained a value of
z$_p=$7.9 for lysozyme charge at infinite dilution, instad of
z$_p=$6.7 received from extrapolated limiting diffusion
coefficients for aqueous lysozyme chloride \cite{2} at pH$=4.5$
and T$=298$ K (these limiting values of the diffusion coefficients
were obtained by extrapolating the diffusion coefficients measured
at low, but nonzero, concentrations of protein). Our value z$_p$
should be compare with z$_p=$3.9 obtained by a Harned-type
analysis \cite{15}, in which the charge in the Debye-H\"{u}ckel
limiting law was adjusted to match the concentration dependence of the difussion data \cite{2}.\\
From ternary lys-salt-water systems conductivities data we
received the folowing values: z$_p=7.12$ from Lys-NaCl-Water,
z$_p<6.42$ from Lys-KCl-Water, and z$_p=5.95$ from
Lys-HN$_4$Cl-Water and we can compare with z$_p=8.9$ obtained at
higher lysozyme concentrations from thermodinamic data for
Lys-NaCl-Water system \cite{14} at pH$=4.5$ and T$=298$ K. \\
The obtained experimental data for binary lys-water, salt-water
and ternary lys-salt-water viscosities, allowed us to calculate
the mobilities of positive ions and to compare the dependence of
the ratio lys/cation mobility ($u_p/u_+$) of the salt
concentrations and to presume its influence on the protein charge
dependence for all three ternary systems (Fig. 5 and Fig. 6). \\ \\

\noindent {\bf Acknowledgment.} One of the authors (DB), is very
grateful to O. Annunziata and J.G. Albright for constant and
helpful advices during the experimental work. This research was
supported by the Texas Christian University Grant RCAF-11950 and
by the NASA Microgravity Biotechnology Program through the Grant
NAG8-1356.

\end{document}